\documentclass[aps,twocolumn,showpacs,preprintnumbers,amssymb]{revtex4}
\def\Box{ }

\def\be{\begin{equation}}
\def\ee{\end{equation}}
\def\bea{\begin{eqnarray}}
\def\eea{\end{eqnarray}}
\def\bma{\begin{mathletters}}
\def\ema{\end{mathletters}}
\def\tr{{\rm tr}}
\def\C{\hbox{$\mit I$\kern-.7em$\mit C$}}

\def\bra#1{\langle#1|} \def\ket#1{|#1\rangle}

\def\proj#1{\ket{#1}\!\bra{#1}}
\def\C{{\cal C}}
\def\B{{\cal B}}
\def\M{{\cal M}}

\usepackage{dcolumn} 
\usepackage{bm} 

\begin{document}

\title{Entanglement cost of mixed states}

\author{G. Vidal$^{1}$, W. D\" ur$^{2}$, and J. I. Cirac$^{3}$}

\address{
$^{1}$Institute for Quantum Information, California Institute of Technology, Pasadena, CA 91125, USA\\
$^2$ Sektion Physik, Ludwig-Maximilians-Universit\"at M\"unchen,
Theresienstr.\ 37, D-80333 M\"unchen, Germany\\
$^3$ Max--Planck Institut f\"ur Quantenoptik, Hans--Kopfermann
Str. 1, D-85748 Garching, Germany}

\date{\today}

\begin{abstract}

We compute the entanglement cost of several families of bipartite mixed states, including arbitrary mixtures of two Bell states. This is achieved by developing a technique that allows us to ascertain the additivity of the entanglement of formation for any state supported on specific subspaces. As a side result, the proof of the irreversibility in asymptotic local manipulations of entanglement is extended to two-qubit systems.

\end{abstract}

\maketitle

Developing a theory of entanglement is considered a priority in
the field of quantum information, where quantum correlations are a
precious resource for information processing \cite{Ni99}. In
particular, the quest for proper entanglement measures has
received much attention in recent years \cite{Ho01}. From the
identification and study of properties of such measures a gain of
insight into the nature of entanglement is expected. In turn,
their computation for particular states provide us with an account
of the resources present in those states.

Two measures of entanglement stand out due to their physical
meaning. Both of them refer to the possibility of transforming
entangled states of a bipartite system by means of local
operations and classical communication (LOCC). The distillable
entanglement \cite{Be96,Ra99}  $E_d(\rho)$ quantifies how much
pure-state entanglement can be extracted from $\rho$. More
specifically, it gives the ratio $M/N$ in the large $N$ limit,
where $M$ is the number of ebits [i.e. entangled bits, or
maximally entangled states $(\ket{00}+\ket{11})/\sqrt{2}$ of a two-qubit system] that can be
distilled from the state $\rho^{\otimes N}$ using LOCC. The
entanglement cost \cite{Be96,Ha00}  $E_c(\rho)$ quantifies, instead, 
the amount of pure--state entanglement needed to create
$\rho$. It is defined in the limit of large $N$ as the ratio
$M/N$, where $M$ is the number of ebits required to prepare
$\rho^{\otimes N}$ using LOCC.

The outputs produced so far by entanglement theory concerning
these two entanglement measures include, among others, the
following remarkable results:

($i$) All forms of bipartite pure-state entanglement are
equivalent in the asymptotic limit \cite{Be96b}, in the sense that
for large $N$ and any bipartite pure state $|\Psi\rangle$,
$|\Psi\rangle^{\otimes N}$ can be {\em reversibly} converted into
ebits. Thus, for pure states $E_d(\ket{\Psi})=E_c(\ket{\Psi})$, with the so--called entropy of entanglement $E(\ket{\Psi})$ denoting the resulting unique measure.

($ii$) Two forms of bipartite entanglement, namely free and bound
entanglement \cite{Ho98}, have been identified for mixed states.
The first form corresponds to mixed states that can be distilled, i.e.
$E_d> 0$. Bound entangled states were defined as those that cannot
be distilled into pure-state entanglement, i.e. $E_d = 0$, in
spite of the fact that they cannot be produced [in the
non-asymptotic regime] by just mixing product (i.e. unentangled)
pure states.

($iii$) Contrary to the pure--state case, the asymptotic
manipulation of some entangled mixed states is {\em irreversible}
\cite{Vi01}. This follows from the gap observed between the
distillable entanglement and the entanglement cost, $E_d < E_c$,
for some mixed states. This phenomenon occurs both for bound
entangled states and for distillable states.

Notice that, as far as mixed states are concerned, the above results are qualitative
\cite{footnote2}. In particular, {\em the entanglement cost $E_c$ has
not been computed for any mixed state}. This problem is related to
the one of the additivity of the entanglement of formation
$E_f(\rho)$ \cite{Be96,Wo01}, an auxiliary measure
that quantifies how much pure--state entanglement ---as
given by $E$--- is required to create
a {\em single} copy of the mixed state $\rho$. In particular, it is not known
whether $E_f(\rho^{\otimes N})=NE_f(\rho)$, which would imply that $E_c=E_f$.

In this paper we compute the value of the entanglement cost $E_c$ for all mixed states $\rho_V$ supported on some specific subspaces $V \subset H_A\otimes H_B$. This is achieved by showing that the entanglement of formation $E_f$ is additive for the tensor product $\rho_V\otimes \sigma$, 
\be
E_f(\rho_V\otimes \sigma) = E_f(\rho_V) + E_f(\sigma),
\label{addit}
\ee
where $\sigma$ is an arbitrary bipartite state, which by iteration implies $E_c(\rho_V)=E_f(\rho_V)$. We also present a technique that allows us to evaluate $E_f$ for some classes of mixed states.

Our considerations include, in a two-qubit system, a  mixture $\rho_p$ of two Bell states, $\ket{\Phi^{\pm}} \equiv (\ket{00} \pm \ket{11})/\sqrt{2}$,
\be
\rho_p \equiv (1-p)\proj{\Phi^+} + p \proj{\Phi^-}, ~~~~p \in [0,\frac{1}{2}], 
\label{mix}
\ee
for which we obtain 
\be
E_c(\rho_p) = H_2(\frac{1}{2} + \sqrt{p(1-p)}),
\label{E_crho}
\ee
$H_2(x)$ being the Shannon entropy $S(x,1-x)$. The distillable entanglement of $\rho_p$ reads \cite{distillable}
\be
E_d(\rho_p) = 1 - H_2(p),
\label{hashing}
\ee
and thus $E_d(\rho_p) < E_c(\rho_p)$ for all $p \in (0,1/2)$. 
That is, even the process of preparing the elementary mixture $\rho_p$ is irreversible, in that not all the pure--state entanglement employed can be subsequently recovered by asymptotic LOCC. This constitutes a new, remarkably simple instance of the irreversibility that takes place in the asymptotic manipulation of entanglement.

Mathematically, the entanglement of formation of a mixed state $\rho\in \B(H_A\otimes H_B)$ can be expressed as \cite{Be96,Wo98}
\be
\label{defEf} E_f(\rho) = \inf_{d\in D_\rho} \sum_k p_k E(\ket{\psi_k}),
\ee
where the entropy of entanglement $E(\ket{\psi})$ is given by $S(\rho_A)$, $\rho_A \equiv \tr_B \proj{\psi}$, and the minimization is performed over the set $D_\rho$ of all pure--state realizations $d \equiv\{p_k, \ket{\psi_k}\}$ of $\rho$, $\rho = \sum_k p_k \proj{\psi_k}$. The entanglement cost, in turn, corresponds to \cite{Be96,Ha00}
 \be
E_c(\rho) = \lim_{N\to \infty} \frac{E_f(\rho^{\otimes N})}{N}.
\ee
Our first goal is to show the additivity of $E_f$ for some mixed states $\rho_V$, as expressed in Eq. (\ref{addit}), which implies that $E_c(\rho_V)=E_f(\rho_V)$. For concreteness, we start by discussing a simple example. After that a theorem announces the result in its full generality. 

{\em Example 1}. Let us consider two qubits $A$ and $B$, with Hilbert spaces $H_A = H_B = \C^2$, and a subspace $V \subset H_A\otimes H_B$ spanned by the vectors \cite{noteTensor}
\bea
\ket{0}_{V} &\equiv& \ket{0}_A \ket{0}_B,\nonumber\\
\ket{1}_{V} &\equiv& \ket{1}_A \ket{1}_B.
\eea
Notice that, in particular, the mixture $\rho_p$ of Eq. (\ref{mix}) is supported on $V$, $\rho_p \in \B(V)$. For any vector $\ket{\phi}_V \in V$, $\ket{\phi}_V = c_0 \ket{0}_V + c_1 \ket{1}_V$, we can define a vector $\ket{\phi}_A \in H_A$ as $\ket{\phi}_A \equiv c_0 \ket{0}_A + c_1 \ket{1}_A$. Then the operation of tracing out qubit $B$ 
\be
\ket{\phi}_V\bra{\phi} \rightarrow tr_B(\ket{\phi}_V\bra{\phi})= \sum_{\alpha=0,1} |c_\alpha|^2 \ket{\alpha}_A\bra{\alpha},
\ee
which is a trace-preserving, completely positive (TPCP) map from $\B(V)$ to $\B(H_A)$, can also be described as a TPCP map $\M$ from $\B(H_A)$ to $\B(H_A)$,
\be
tr_B(\ket{\phi}_V\bra{\phi})=\M(\ket{\phi}_A\bra{\phi}),
\ee
given by
\be
\M(X) = \sum_{\alpha=0,1} \tr(\ket{\alpha}_A\bra{\alpha}X)~~\ket{\alpha}_A\bra{\alpha}.
\label{M1}
\ee
The relevant feature of subspace $V$ is that $\M$ is an {\em entanglement--breaking} map \cite{Sh02}, as Eq. (\ref{M1}) makes manifest \cite{entbreak}. The theorem below establishes that this property alone guarantees that Eq. (\ref{addit}) holds for any $\rho_V \in \B(V)$, and thus $E_c(\rho_V) = E_f(\rho_V)$. Then we can use the closed expression for $E_f$ in two-qubit systems \cite{Wo98} to evaluate $E_f(\rho_V)$, and thereby obtain, for instance, the value of $E_c(\rho_p)$ displayed in Eq. (\ref{E_crho}).

More generally, the theorem considers four arbitrary quantum systems, denoted $A$, $B$, $a$ and $b$, and refers to the entanglement between $Aa$ and $Bb$. $V$ is a subspace of $H_{AB}$ such that, for any $\ket{\Psi}_{Vab} \in V\otimes H_{ab}$, tracing out subsystem $B$ destroys all existing entanglement between $AB$ with $ab$,
\be
\tr_B (\ket{\Psi}_{Vab}\bra{\Psi}) = \sum_l q_l \ket{\mu_l}_A\bra{\mu_l}\otimes \ket{\nu_l}_{ab}\bra{\nu_l},
\label{sepa}
\ee
that is, such that the map $\B(V)\rightarrow \B(H_A)$ given by $\rho_V \rightarrow \tr_B \rho_V$ is entanglement--breaking \cite{Sh02}.

\vspace{1mm}

{\bf Theorem:} Let $\rho_V \in \B(V)$ and $\sigma_{ab} \in \B(H_{ab})$. Then
\be
E_f(\rho_V\otimes\sigma_{ab}) = E_f(\rho_V) +  E_f(\sigma_{ab}).
\ee

\vspace{1mm}

{\em Proof:} Notice that $E_f(\rho_V\otimes\sigma_{ab}) \leq E_f(\rho_V) +  E_f(\sigma_{ab})$, because from optimal pure--state decompositions of $\rho_V$ and of $\sigma_{ab}$ a (possibly non-optimal) decomposition for $\rho_V\otimes\sigma_{ab}$ can be constructed with average $E$ given by $E_f(\rho_V) +  E_f(\sigma_{ab})$. In what follows we will show that
\be
E_f(\rho_V\otimes\sigma_{ab}) \geq E_f(\rho_V) +  E_f(\sigma_{ab}).
\label{desi}
\ee
Let us consider a decomposition $\{p_k, \ket{\Psi_k}\}$ of $\rho_V\otimes\sigma_{ab}$,
\be
\rho_V\otimes\sigma_{ab} = \sum_k p_k \ket{\Psi_k}\bra{\Psi_k},
\ee
such that it is optimal, that is,
\be
E_f(\rho_V\otimes\sigma_{ab}) = \sum_k p_k E(\ket{\Psi_k}). 
\ee
We recall that all $\ket{\Psi_k}$ must belong to $V \otimes H_{ab}$. Next we argue that in order to prove Eq. (\ref{desi}) ---and therefore the theorem--- it is sufficient to show that for any pure state $\ket{\Psi}_{Vab} \in V\otimes H_{ab}$, 
\be
E(\ket{\Psi}) \geq E_f(\varrho_{V}) + E_f(\pi_{ab}),
\label{desigualtat}
\ee
where 
\bea
\varrho_{V} &\equiv& \tr_{ab}(\ket{\Psi}_{Vab}\bra{\Psi}), \\
\pi_{ab} &\equiv& \tr_{AB}(\ket{\Psi}_{Vab}\bra{\Psi}).
\eea
Indeed, denoting by $\varrho^k_{V}$ and $\pi^k_{ab}$ the reduced density matrices of systems  $AB$ and $ab$ for each $\ket{\Psi_k}$, we would have 
\bea
E_f(\rho_V\otimes\sigma_{ab}) &=& \sum_k p_k E(\ket{\Psi_k}) \nonumber\\
&\geq& \sum_k p_k  E_f(\varrho^k_{V}) +  \sum_k p_k E_f(\pi^k_{ab}) \nonumber\\
&\geq&  E_f(\sum_k p_k \varrho^k_{V}) +  E_f(\sum_k p_k\pi^k_{ab}) \nonumber\\
&=& E_f(\rho_V) +  E_f(\sigma_{ab}).
\eea
where the first inequality assumes Eq. (\ref{desigualtat}), the second inequality uses that $E_f$ is a convex function and the last step follows from the fact that $\rho_V=\sum_k p_k \varrho^k_V$ and $\sigma_{ab} = \sum_k p_k \pi_{ab}^k$. Let us then move to justify Eq. (\ref{desigualtat}). $E(\ket{\Psi}_{Vab})$ is given by the von Neumann entropy of the reduced density matrix $\xi_{Aa} \equiv \tr_{Bb}(\ket{\Psi}_{Vab}\bra{\Psi})$. Because of Eq. (\ref{sepa}) we have
\bea
\xi_{Aa} &=& \sum_l q_l \ket{\mu_l}_A\bra{\mu_l}\otimes \tr_b(\ket{\nu_l}_{ab}\bra{\nu_l}) \nonumber \\
&\equiv& \sum_l q_l \ket{\mu_l}_A\bra{\mu_l} \otimes \eta_a^l.
\eea
Define $\gamma_A \equiv \tr_a (\xi_{Aa}) = \sum_l q_l \ket{\mu_l}_A\bra{\mu_l}$. Then we have
\bea
E(\ket{\Psi}_{Vab}) = S(\xi_{Aa})
&\geq& S(\gamma_A) + \sum_l q_l S(\eta^l_a) \label{strong}\nonumber\\
&\geq& E_f(\varrho_V) + \sum_l q_l E(\ket{\nu_l}_{ab})\nonumber\\
&\geq& E_f(\varrho_V) + E_f(\pi_{ab}).
\eea
The first inequality follows from the strong subadditivity of the entropy \cite{notestrong}, as shown in \cite{Sh02}. In the second inequality we have used that $ S(\gamma_A = \tr_B(\varrho_V)) \geq E_f(\varrho_V)$ \cite{notebound}, also that $S(\eta^l_a = \tr_{b} \ket{\nu_l}_{ab}\bra{\nu_l}) = E(\ket{\nu_l}_{ab})$. Finally, the last inequality follows from the fact that $\{q_l,\ket{\nu_l}_{ab}\}$ is a (possibly non--optimal) realization of $\pi_{ab}$, $\pi_{ab} = \sum_l q_l  \ket{\nu_l}_{ab}\bra{\nu_l}$. $\Box$

\vspace{1mm}

Thus, as illustrated in example 1, we can use this theorem to relate the asymptotic entanglement cost $E_c$ of the mixed states supported on some subspace $V\in H_A\otimes H_B$ to their entanglement of formation $E_f$. All that is needed is to identify subspaces $V\subset H_{AB}$ that fulfill the above requirements.

Recall that, other than for two--qubit mixed states, the value of $E_f$ is only known in very few cases \cite{Wo98}. In this sense, another class of subspaces $V' \in H_A\otimes H_B$ of particular interest are those such that all their vectors are related by local unitary transformations. Since their reduced density matrices have the same spectrum, all these states are equally entangled. Let $E(V')$ denote their entropy of entanglement. Then, because a mixed state $\rho_{V'}$ supported on $V'$ is necessarily a mixture of vectors of $V'$, we conclude that $E_f(\rho_{V'})=E(V')$. 

{\em Example 2}. Let us consider a two-qutrit system, $H_A=H_B=\C^3$, and the {\em antisymmetric} subspace $V'\in H_A\otimes H_B$, spanned by the vectors
\bea
\ket{0}_{V'}&\equiv&\frac{1}{\sqrt{2}}(\ket{1}_A\ket{2}_B - \ket{2}_A\ket{1}_B),\nonumber\\
\ket{1}_{V'}&\equiv&\frac{1}{\sqrt{2}}(\ket{2}_A\ket{0}_B - \ket{0}_A\ket{2}_B),\nonumber\\
\ket{2}_{V'}&\equiv&\frac{1}{\sqrt{2}}(\ket{0}_A\ket{1}_B - \ket{1}_A\ket{0}_B).
\eea

Notice that $\tr_B (\ket{\alpha}_{V'}\bra{\beta}) = (\delta_{\alpha,\beta}I_A - \ket{\beta}_A\bra{\alpha})/2$. Therefore, for any vector $\ket{\phi}_{V'} = \sum_{\alpha} c_\alpha \ket{\alpha}_{V'}$, its reduced density matrix $\rho_A^{\phi}\equiv \tr_B\ket{\phi}_{V'}\bra{\phi}$ can be expressed as a linear combination of the identity operator $I_A$ and the transposition of a projector onto the state $\ket{\phi}_A \equiv \sum_\alpha c_{\alpha} \ket{\alpha}_A$,
\be
\rho_A^{\phi} = \frac{1}{2}I_A - \frac{1}{2}(\ket{\phi}_A\bra{\phi})^T
\ee
(equivalently, any two antisymmetric states are related by local unitary transformations). It is then straightforward to obtain $E_f(\rho_{V'}) = 1$ ebit for any state $\rho_{V'}\in \B(V')$. Unfortunately, the subspace $V'$ does not meet the requirements of the theorem, and we can not compute $E_c$ for this family of mixed states.

Finally, in some cases one can combine the two results discussed in this paper to determine both $E_f$ and $E_c$, as the following examples illustrate.

{\em Example 3}. Let us consider a qubit-qutrit system, $H_A=\C^2$ and $H_B = \C^3$, and the subspace $V''$ spanned by
\bea
\ket{0}_{V''} \equiv \frac{1}{\sqrt{3}}(\ket{0}_A\ket{2}_B - \sqrt{2}\ket{1}_A\ket{0}_B), \nonumber\\
\ket{1}_{V''} \equiv \frac{-1}{\sqrt{3}}(\ket{1}_A\ket{2}_B - \sqrt{2}\ket{0}_A\ket{1}_B).
\eea
In this case $\tr_{B} (\ket{\alpha}_{V''}\bra{\beta}) = (2\delta_{\alpha,\beta}I_A - \ket{\alpha}_A \bra{\beta} )/3$, and therefore, for any vector $\ket{\phi}_{V''} = \sum_{\alpha} c_\alpha \ket{\alpha}_{V''}$, we find
\be
\tr_B(\ket{\phi}_{V''}\bra{\phi}) = \frac{2}{3}I_A -  \frac{1}{3}\ket{\phi}_A\bra{\phi},
\label{traca3}
\ee
where $\ket{\phi}_A \equiv \sum_{\alpha} c_\alpha \ket{\alpha}_{A}$. It follows that entanglement is constant in $V''$, $E_f(\rho_{V''}) = E(\ket{\phi}_{V''}) = H_2(1/3)$. In addition, by noticing that the TPCP map $\M''$,
\be
\M''(\ket{\phi}_A\bra{\phi}) \equiv \frac{2}{3}I_A - \frac{1}{3} \ket{\phi}_A \bra{\phi},
\ee
is entanglement--breaking, since it can be expanded as \cite{entbreak}
\be
\M''(X) = \int_{H_A} d\ket{\phi}_A \tr(\ket{\phi}_A \bra{\phi} X) ~~(I_A-\ket{\phi}_A \bra{\phi}),
\ee
it follows, because of the theorem, that also $E_c(\rho_{V''}) = H_2(1/3)$.

So far, we have calculated the entanglement cost for rank 2 density operators. 
Obviously, we can use our methods to determine this quantity for higher rank 
operators, as the following example shows.

{\em Example 4}. Let us consider $H_A=\C^3$, $H_B=\C^6$, and the subspace $V'''$ spanned by 
\bea
\ket{0}_{V'''}\!\!&\equiv&\frac{1}{2}(\ket{1}_A\ket{2}_B + \ket{2}_A\ket{1}_B + \sqrt{2}\ket{0}_A\ket{3}_B), \nonumber\\
\ket{1}_{V'''}\!\!&\equiv&\frac{1}{2}(\ket{2}_A\ket{0}_B + \ket{0}_A\ket{2}_B + \sqrt{2}\ket{1}_A\ket{4}_B),\nonumber\\
\ket{2}_{V'''}\!\!&\equiv&\frac{1}{2}(\ket{0}_A\ket{1}_B + \ket{1}_A\ket{0}_B + \sqrt{2}\ket{0}_A\ket{5}_B).
\eea

Since $\tr_{B} (\ket{\alpha}_{V'''}\bra{\beta}) = (\delta_{\alpha,\beta}I_A + \ket{\beta}_A \bra{\alpha})/4$, for any vector $\ket{\phi}_{V'''} = \sum_{\alpha} c_\alpha \ket{\alpha}_{V'''}$ we have
\be
\tr_B(\ket{\phi}_{V'''}\bra{\phi}) = \frac{1}{4}I_A + \frac{1}{4}(\ket{\phi}_A\bra{\phi})^T,
\label{traca4}
\ee
where $\ket{\phi}_A \equiv \sum_{\alpha} c_\alpha \ket{\alpha}_{A}$. It follows that the spectrum of the reduced density matrix for system $A$ is $\{1/2,1/4,1/4\}$, that is, the same for any pure state $\ket{\phi}_{V'''}$. Therefore the entanglement is also constant in $V'''$, $E_f(\rho_{V'''}) = E(\ket{\phi}_{V'''}) = 1.5$ ebits. Finally, the TPCP map
\be
\M'''(\ket{\phi}_A\bra{\phi}) \equiv \frac{1}{4}I_A  + \frac{1}{4}(\ket{\phi}_A\bra{\phi})^T,
\ee
is entanglement--breaking \cite{alternative}, since when applied to the maximally entangled state $(\sum_{i=0}^2 \ket{i}_A\ket{i}_C)/\sqrt{3}$, where $C$ denotes an auxiliary system, the resulting state $P_+/6$, proportional to the projector $P_+$ onto the symmetric subspace of $H_A\otimes H_C$, is known to be separable \cite{Du00}. Consequently, the theorem implies that $E_c(\rho_{V'''})=E_f(\rho_{V'''})=1.5$ ebits.

Summarizing, we have shown that the entanglement of formation $E_f$ is additive for mixed states supported on a subspace such that tracing out one of the parties corresponds to an entanglement--breaking channel \cite{Sh02}. This has allowed us to evaluate the entanglement cost $E_c$ of several families of mixed states. A series of examples have been selected to illustrate these results.

Whether the entanglement of formation is additive for general mixed states remains an open question, which certainly deserves further investigation.


This work was supported by European Community,  under project EQUIP
(contract IST-1999-11053), grants HPMF-CT-1999-00200 (G.V.) and HPMF-CT-2001-01209 (W.D.) [Marie Curie fellowships] and through the ESF; by the Institute for Quantum Information GmbH, and by the United States of America through the NSF, under Grant. No. EIA-0086038.



\begin{references}
\bibitem{distillable} Eq. (\ref{hashing}) is the distillation ratio obtained through the hashing protocol \cite{Be96},  subsequently shown to be optimal in \cite{Ra97}.
\bibitem{noteTensor}
Throughout this paper we will omit tensor products if there is no danger of confusion.
\bibitem{entbreak} As discussed in \cite{Sh02}, a TPCP map $\M$ is entanglement--breaking if and only if it can be written as $\M(X) = \sum_i \tr(M_iX)\rho_i$, where the positive operators $M_i$ add to the identity, $\sum_i M_i =I$, and $\rho_i$ are density matrices.
\bibitem{notestrong} More precisely, inequality of Eq. (\ref{strong}) follows from applying the strong superadditivity of the von Neumann entropy, in the form $S(\sigma_{Aa}) \ge S(\sigma_A) + S(\sigma_{AaC}) - S(\sigma_{AC})$, to the state $\sum_l q_l \ket{\mu_l}_{A}\bra{\mu_l}\otimes \eta_a^l \otimes \ket{l}_C\bra{l}$, where $\{\ket{l}_C\}$ is an orthonormal set of vectors.
\bibitem{notebound} Recall that for any mixed state $\rho_{CD}$ and any of its realizations $\{p_k,\phi_k\}$,  the concavity of the von Neumann entropy implies that
\bea
E_f(\rho_{CD}) \leq \sum_k p_k E(\phi_k) = \sum_k p_k S\left(\tr_D[\ket{\phi_k}_{CD}\bra{\phi_k}]\right) \\
\leq S\left(\sum_k p_k\tr_D[\ket{\phi_k}_{CD}\bra{\phi_k}]\right) = S(\tr_D[\rho_{CD}]).
\eea
\bibitem{alternative} An alternative characterization of an entanglement--breaking channel $\M:\B(H)\rightarrow\B(H)$ is that it maps a maximally entangled state $(\sum_{i=1}^d \ket{i}\ket{i})/\sqrt{d} \in H\otimes H_C$, where $H_C$ is an auxiliary system and $d = \dim H= \dim H_C$, into a separable state, following the ideas of \cite{Ja72}.
\end{references}
\end{document}